\documentclass[english]{cccconf}
\usepackage[comma,numbers,square,sort&compress]{natbib}
\usepackage{epstopdf}
\usepackage{mathrsfs}
\usepackage{amsfonts}
\usepackage{amssymb}
\usepackage{amsmath}
\usepackage{cases}
\usepackage{indentfirst}
\usepackage{subfigure}
\usepackage{booktabs}
\usepackage{pifont}
\usepackage{graphicx}
\usepackage{lmodern}
\usepackage{threeparttable}
\usepackage[ruled,linesnumbered]{algorithm2e}
\usepackage{algorithmic}
\begin{document}

\title{Appointed-Time Fault-Tolerant Control for Flexible Hypersonic Vehicles with Unmeasurable States Independent of Initial Errors}

\author{Tianlong Zhao,
        Fei Hao}



\affiliation[]{The Seventh Research Division, School of Automation Science and Electrical Engineering, Beihang University, Beijing, 100191, China.
        \email{ztl@buaa.edu.cn};fhao@buaa.edu.cn}

\maketitle

\begin{abstract}This article aims to derive a practical tracking  control algorithm for flexible air-breathing hypersonic vehicles (FAHVs) with lumped disturbances, unmeasurable states and actuator failures. Based on the framework of the backstepping technique, an appointed-time fault-tolerant protocol independent of initial errors is proposed. Firstly, a new type of a state observer is constructed to reconstruct the unmeasurable states. Then, an error transformation function is designed to achieve prescribed performance control that does not depend on the initial tracking error. To deal with the actuator failures, practical fixed-time neural network observers are established to provide the estimation of the lumped disturbances. Finally, the proposed control strategy can ensure the practical fixed-time convergence of the closed-loop system, thereby greatly enhancing the transient performance. The proposed method addresses the challenges of ensuring real-time measurement accuracy for angle of attack and flight path angle in hypersonic vehicles, coupled with potential sudden actuator failures, effectively overcoming the drawback of prescribed performance control that requires knowledge of initial tracking errors. Some simulation results are provided to demonstrate the feasibility and the effectiveness of the proposed strategy. 
\end{abstract}


\footnotetext{This work is supported by National Natural Science
Foundation of China under Grant 61573036.}

\section{Introduction}
For hypersonic cruise vehicles, transient performance is a critically important metric of their control systems\cite{label1}. However, hypersonic vehicles are inherently strong coupled, nonlinear multivariable systems, which present significant challenges for trajectory tracking control. To address these challenges, various control methods have been developed, such as neural network control, fixed-time and finite-time strategies\cite{label5}. While these control methods can achieve good tracking of command signals and provide a certain level of robustness, they often cannot simultaneously optimize both the steady-state and transient performance.

To overcome this limitation, modified versions of PPC for hypersonic vehicles have been proposed, aiming to design new prescribed performance functions that eliminate dependence on initial errors\cite{label7}. Unfortunately, these methods may lead to excessive overshoot because the initial value of the performance function needs to be sufficiently large. Moreover, traditional PPC requires the performance function to be steep enough to ensure good transient performance, which can cause the system to stay very close to the performance boundary during transients, resulting in overly large control inputs and deep saturation of the controller, potentially causing loss of control of the hypersonic vehicle.

In response to the issues currently faced by trajectory tracking control for hypersonic vehicles, a novel appointed-Time Fault-Tolerant Control algorithm independent of initial error values is designed. Considering practical problems such as some states being difficult to measure and potential actuator failures during actual flights of hypersonic vehicles, new differential trackers and fault-tolerant controllers are designed.

The key contributions of this paper are outlined as follows. Firstly, a practical fixed-time neural network observer is designed to achieve rapid and accurate observation of unmodeled dynamics, aerodynamic parameter uncertainties, external disturbances, and actuator faults. Then, by introducing prescribed performance control and practical fixed-time controllers, the hypersonic vehicle can achieve satisfactory transient and steady-state performance while reducing control gains and making control signals smoother. Additionally, a novel error transformation function is introduced so that when the initial tracking error is unknown, the transformed error remains within the designed performance boundaries. Finally, to solve the problem of certain states being difficult to measure in hypersonic vehicles, a second-order nonlinear differential tracker is introduced, which facilitates the practical application of the controller.

\section{Problem Formulation}

\subsection{The longitudinal dynamics of FAHV}
Referring to \cite{label9}, the longitudinal dynamics of the FAHV can be structured in the following form:
\begin{equation}
\left\{ {\begin{aligned}
		&{\dot V = {g_V}\Phi  + {f_V} + {d_V}}\\
		&{\dot h \approx {g_h}\gamma \;}\\
		&{\;\dot \gamma  = {g_\gamma }\theta  + {f_\gamma } + \;{d_\gamma }}\\
		&{\;\dot \theta  = {g_\theta }Q + {f_\theta } + {d_\theta }}\\
		&{\dot Q = {g_Q}{\delta _e} + {f_Q} + {d_Q}}\\
		&{{{\ddot \eta }_i} =  - 2{\zeta _i}{\omega _i}{{\dot \eta }_i} - \omega _i^2{\eta _i} + {N_i},\;\;i = 1,2,3}
\end{aligned}} \right.
\end{equation}

The hypersonic vehicle is composed of five states $ x ={[V,h,\gamma ,\theta ,Q]^T},$ six flexible states $\eta  = {[{\eta _1},{\dot \eta _1},{\eta _2},{\dot \eta _2},{\eta _3},{\dot \eta _3}]^T}$
and two control inputs $u = {\left[ {{\rm{\Phi }}\;,{\delta _e}} \right]^T}$. $V,h,\gamma ,\theta ,Q$ represent the velocity, altitude, flight path angle, pitch angle and pitch rate, respectively. $\rm{\Phi }, {\delta _e}$ represent the fuel equivalence ratio and deflection of elevator, respectively. For $i=V,h,\gamma ,\theta ,Q$, $g_i$ and $f_i$ are both constituted by aerodynamic coefficients. $d_i, i=V,h,\gamma ,\theta ,Q$ indicate the lumped disturbances that encompass unmodeled dynamics, uncertainties in aerodynamic parameters, and external perturbations. $\eta _i, i=1,2,3$ are three different frequency flexible states on the fuselage and $\zeta_i, \omega _i, N_i$ represent their damping ratio, natural frequency and generalized forces, respectively. 
 
\noindent\textbf{Assumption 1.} $g_i \neq 0, i=V,h,\gamma ,\theta ,Q$ hold throughout the entire cruising envelope of FAHV.

\noindent\textbf{Assumption 2.} According to the engineering practice, $d_i$ and their derivatives are bounded.

\noindent\textbf{Assumption 3.} During the cruise phase, the range of system state changes is relatively small, therefore the speed of system state changes is much lower than that of lumped disturbances.

\noindent\textbf{Lemma 1.}\cite{label10} Consider the system $\dot x = f(t,x),x(0) = {x_0}$. If there exists a Lyapunov function $V(x)$ to satisfy:

\[\dot V(x) \le  - {c_1}{V^p}(x) - {c_2}{V^q}(x) + \varsigma, \]
where ${c_1},{c_2} > 0,p \in (0,1),q \in (1,\infty ),\varsigma  > 0$, the origin of the system $x(t)$ is practical fixed-time stable. Further, the residual set of the solution can be estimated using the following inequality:
\[x \in \{ V(x) \le \min \{ {(\frac{\varsigma }{{(1 - w){c_1}}})^{\frac{1}{p}}},{(\frac{\varsigma }{{(1 - w){c_2}}})^{\frac{1}{q}}}\} \},\]
where $0 < w < 1$ being a constant, the convergence time $T$ satisfies:
\[T \le \frac{1}{{{c_1}w(1 - p)}} + \frac{1}{{{c_2}w(q - 1)}}.\]
\noindent\textbf{Lemma 2.}\cite{label10} For any ${x_1},{x_2},...,{x_n} \in R$, one has
\[{(\sum\limits_{i = 1}^n {|{x_i}|} )^\alpha } \le \max \{ {n^{\alpha  - 1}},1\} (\sum\limits_{i = 1}^n {|{x_i}{|^\alpha }} ),\]
with $\alpha $ is a positive constant.

\noindent\textbf{Lemma 3.}\cite{label8} For any $x \in R$, one has
\[0 < |x| - x\tanh (\frac{x}{l}) \le cl,\]
where $c = 0.2785, l$ is a positive constant.

\noindent\textbf{Lemma 4.}\cite{label11} For any continuous function $f(x):\Omega \to R$, where $\Omega \subset R^n$ is a compact set, we have $\hat f(x) = {{\hat W}^T}h(x)$,where ${\hat W} \subset R^q$ are the weights of neural network and $x \subset R^n$ are the input states of neural network, where $h(x) = {[{h_1},...,{h_q}]^T},{h_i} = \exp ( - {(x - {\delta _i})^T}(x - {\delta _i})/{{\hat b}^2})$,$\delta _i$ is the network center of Gaussian function at $i$th node and $\hat b$ is the width of Gaussian function at $i$th node. For any given constant $\epsilon_N>0$ and by appropriately choosing $\hat b$ and $\delta_i$, and with sufficiently large $q$, there always exists an RBF (Radial Basis Function) neural network ${W^{*T}}h(x)$ such that $f(x) = {W^{*T}}h(x) + \varepsilon (x),|\varepsilon (x)| < {\varepsilon _N},\forall x \in {\Omega _x}$.

\subsection{Observation of a Longitudinal Model with Unknown States}
During the cruise of a hypersonic vehicle, the flight path angle and angle of attack are typically small, making it difficult for sensors to obtain precise measurements of these states. Therefore, we need to estimate these states using observable known states $V,h,Q,\theta$. From the model of the hypersonic vehicle, it can be seen that to reconstruct the unknown state 
$\gamma$, information from state $\dot h$ is required, i.e., $\gamma  = \arcsin (\frac{{\dot h}}{V})$. To address this, an observer has been designed to reconstruct the states $h$ and $\dot h$:
\begin{equation}
	\left\{ {\begin{array}{*{20}{l}}
			{\dot {\hat h} = {\chi _h}}\\
			{{\dot{{\chi }_h}} =  - {d_h}^2[\mathrm{sig}(\hat h - h;{\eta _0},{\eta _1}) + \mathrm{sig}(\frac{{{\chi _h}}}{{{d_h}}};{\eta _2},{\eta _3})],}
	\end{array}} \right.
\end{equation}
where $\mathrm{sig}(\hat h - h;{\eta _0},{\eta _1}) = {\eta _0}[{(1 + {e^{ - {\eta _1}(\hat h - h)}})^{ - 1}} - 0.5]$, based on Lemma 1 derived from \cite{label12}, we can conclude that $\hat h$ converges to $h$ and $\dot{\hat h}$ converges to ${\dot h}$. Therefore, $\hat \gamma  = \arcsin (\frac{{{\chi _h}}}{V})$, and further, from $\alpha  = \theta  - \gamma $, it follows that $\hat \alpha  = \int {Qdt}  - \hat \gamma $. To avoid computing the derivative of the angle of attack $\alpha$, the pitch angle $\theta$ was utilized as a state. Furthermore, to reconstruct the state equations of the system, an observer has been designed to estimate states $\hat \gamma$ and $\dot{\hat \gamma}$:
\begin{equation}
	\left\{ {\begin{array}{*{20}{l}}
			{{{\dot s}_1} = {s_2}}\\
			{{{\dot s}_2} =  - {d^2}[\mathrm{sig}({s_1} - \hat \gamma ;{\eta _0},{\eta _1}) + \mathrm{sig}(\frac{{{s_2}}}{d};{\eta _2},{\eta _3})],}
	\end{array}} \right.
\end{equation}

By designing the observer to make $s_1$ converge to $\hat{\gamma}$, we establish a cascaded observation mechanism. This allows $s_2$ to effectively estimate $\dot{\gamma}$ through its convergence to $\dot{\hat{\gamma}}$, which in turn converges to $\dot{\gamma}$.

\subsection{Actuator fault models}
The following actuator fault models are established: ${\delta _e} = {\lambda _\delta }{\delta _{ed}} + {f_\delta }, \Phi  = {\lambda _\Phi }{\Phi _d} + {f_\Phi }.$ ${f_i}, i = \delta, \Phi $ represents the unknown constant of the actuator bias fault, ${\lambda _i} \in (0,1], i = \delta, \Phi $ indicates the remaining control capability of the actuator after a fault occurs, and ${\delta _e}, \Phi $ denotes the control signal.

Taking into account the actuator faults and the unknown states, the longitudinal model of the system is transformed into the following form:

\begin{equation}
	\left\{ {\begin{aligned}
			&{\dot V = {g_V}{\Phi _d} + {f_V} + {D_V}}\\
			&{\dot h = {g_h}\hat \gamma  + {D_h}}\\
			&{\dot{\hat{\gamma}}  = {g_\gamma }\theta  + {f_\gamma } + {D_{\hat \gamma }}}\\
			&{\hat \alpha  = \theta  - \hat \gamma }\\
			&{\dot \theta  = {g_\theta }Q + {f_\theta } + {D_\theta }}\\
			&{\dot Q = {g_Q}{\delta _{ed}} + {f_Q} + {D_Q},}
	\end{aligned}} \right.
\end{equation}
where ${D_V} = {g_V}{f_\Phi } + {D_{V1}} + {g_V}{\lambda _\Phi }{\Phi _d} - {g_V}{\Phi _d},{D_Q} = {g_Q}{f_\delta } + {D_{Q1}} + {g_Q}{\lambda _\delta }{\delta _{ed}} - {g_Q}{\delta _{ed}},{D_{V1}} = {d_V} + {\Delta _1},{D_h} = {\Delta _2},{{D_{\hat \gamma }} = {d_\gamma } + {\Delta _3}},{D_\theta } = {d_\theta } + {\Delta _4}$
		$,{{D_{Q1}} = {d_Q} + {\Delta _5}.}$
$\Delta_i, i=1,2,3,4,5$ represents the deviation in aerodynamic parameters and the states themselves caused by the unobservable states 
$\gamma, \alpha$, as observed from $\hat \gamma,\hat \alpha$.

\subsection{Error transformation function}
To overcome the limitations of conventional prescribed performance control, the error transformation function has been designed as follows:
\begin{equation}
\varphi (t) = \left\{ {\begin{array}{*{20}{l}}
		{1 - (1 - \beta ){{(\frac{{2({T_p} - t)}}{{{T_p}({e^{\frac{{\mu t}}{{{T_p}}}}} + 1) - t}})}^\alpha },t \le {T_p}}\\
		{1,t > {T_p}.}
\end{array}} \right.
\end{equation}

The initial tracking error is transformed using the error transformation function. Here, $0 < \beta  < 1$ is a variable to be designed, which is related to the system's convergence rate and the initial error. $a > 1,\mu  > 0$ can adjust the curvature of the function. When $t = 0$, $\varphi (0) = \beta $, with an appropriately chosen $\beta$, the initial error can be mapped to a small number that tends towards zero. After time 
${T_p}$, $\varphi (t) = 1$, and the error transformation function exits the control process.

\subsection{Practical Fixed-Time Neural Network Observer}
A practical fixed-time neural network observer has been designed to approximate the system's lumped errors and actuator faults, with the  expression shown in the following formula:
 \begin{equation}
 	\left\{
 	 \begin{aligned}
 			&{{\dot z}_i} = {{\hat d}_i} +f_i(z_i-x_i)+ {f_i} + {g_i}{{\bar x}_i}\\
 			 &f_i(z_i \!-x_i) \!= \!- {l_{i1}}{{\left\langle {{z_i} \!- {x_i}} \right\rangle }^{{\alpha _1}}} \!- {l_{i2}}{{\left\langle {{z_i} \!- {x_i}} \right\rangle }^{{\beta _1}}} \!- {l_{i3}}({z_i} \!- {x_i})\\
			&{\dot{\hat{W_i}}} = {\Gamma _{wi}}(({z_i} - {x_i})\phi (x) - k{{\hat W}_i})\\
			&{{\hat d}_i} = {{\hat W}_i}^T\phi (x),
 	\end{aligned}
 	\right.
\end{equation}
where ${z_i}$ represents the state of the observer, ${x_i}$ denotes the observed state variables, and the symbol ${\left\langle x \right\rangle ^\alpha }$ stands for ${x^\alpha }{\mathop{\rm sgn}} (x)$. $\hat d$ is used to estimate the lumped disturbances $d$, employing a neural network approximation, where the weights ${\hat W_i}$ are updated based on the observation error of the states.

\noindent\textbf{Theorem 1.} By selecting appropriate gains ${l_{i1}},{l_{i2}},{l_{i3}},k$,
${\Gamma _{wi}}$, and 
${\alpha _1} \in (0,1),{\beta _1} \in (1, + \infty )$, it can be ensured that the state $\hat{d_i}$ will converge to a small neighborhood around $d_i$ within a fixed time.

\noindent\textbf{Proof:} From Lemma 4, we can conclude that ${d_i} = {W^{*T}_i}\phi ({x_i}) + {\varepsilon _i}$. Define the state estimation error, disturbance estimation error, and weight estimation error:
${e_1} = {z_i} - {x_i}$,${e_2} = {{\hat d}_i} - {d_i} = {{\tilde W}^T}_i\phi ({x_i}) - {\varepsilon _i}$,${{\tilde W}_i} = {{\hat W}_i} - W_i^*$.

Define the Lyapunov function:
\begin{equation}
	V = {V_1} + {V_2} = \frac{1}{2}{e_1}^2 + \frac{1}{2}{\tilde W^T}_i\Gamma _w^{ - 1}{\tilde W_i}.
\end{equation}

Differentiation $V$ yields:
\begin{equation}
	\begin{aligned}
		\dot V &=  - {l_{i1}}{e_1}^{{\alpha _1} + 1} - {l_{i2}}{e_1}^{{\beta _1} + 1} - {l_{i3}}{e_1}^2 - {e_1}{\varepsilon _i} - k{{\tilde W}^T}_i{{\hat W}_i}\\
		&\le  - {l_{i1}}{V_1}^{\frac{{{\alpha _1} + 1}}{2}} - {l_{i2}}{V_1}^{\frac{{{\beta _1} + 1}}{2}} - ({l_{i3}} + \frac{1}{2}){e_1}^2  - k{\Gamma _w}{V_2}\\
		&\quad+ {\nu _1} + {\nu _2},
	\end{aligned}
\end{equation}
where ${\nu _1} = \frac{1}{2}{\varepsilon _i}^2, {\nu _2} = \frac{k}{2}W_i^{*T}W_i^*$.
Therefore, all states of the observer are convergent, and furthermore, we can obtain:
\begin{equation}
	{\dot V_1} \le  - {l_{i1}}{V_1}^{\frac{{{\alpha _1} + 1}}{2}} - {l_{i2}}{V_1}^{\frac{{{\beta _1} + 1}}{2}} + \nu,
\end{equation}
where $\nu$ is a bounded variable related to the approximation error of the neural network. From Lemma 1, we can obtain that $z_i$ can converge to a neighborhood of $x_i$ within a fixed time $T$:
{\small
\begin{equation}
	\begin{aligned}
		&{e_1} \in \{ {V_1}({e_1}) \le \min \{ {(\frac{{{\nu _1}}}{{(1 - w){l_{i1}}}})^{\frac{2}{{{\alpha _1} + 1}}}},{(\frac{{{\nu _1}}}{{(1 - w){l_{i2}}}})^{\frac{2}{{{\beta _1} + 1}}}}\} \}\\
		&T \le \frac{1}{{{c_1}w(1 - \frac{{{\alpha _1} + 1}}{2})}} + \frac{1}{{{c_2}w{(^{\frac{{{\beta _1} + 1}}{2}}} - 1)}}.
	\end{aligned}
\end{equation}
}

After time $T$, we have ${z_i} = {x_i} + o$, where $o$ is a sufficiently small parameter. According to Theorem 2 from \cite{label13} and Assumption 3, it follows that ${\dot z_i} \approx {\dot x_i}$. Substituting this into the expression for the observer yields ${e_2} \approx {l_{i1}}{o^{{\alpha _1}}} + {l_{i2}}{o^{{\beta _1}}} + {l_{i3}}o \le {l_{i1}} + ({l_{i2}} + {l_{i3}})o$. Therefore, ${\hat d_i}$ can converge to a neighborhood of $d_i$ at time $T$. with the convergence residual being proportional to the observer's convergence residual. Then we can define the observation error 
${E_i} = {d_i} - {\hat d_i},i = V,r,\theta ,Q$, and ${E_i}$ is bounded.

\section{Main Results}
According to \cite{label14}, we can decompose the equation (4) into two distinct components: velocity subsystem and altitude subsystem. In the following sections, we will detail the design of the controllers for each of these subsystems, ensuring that they can operate effectively to maintain precise control over velocity and altitude independently.
\subsection{Velocity Controller Design:}
Define the velocity tracking error as ${e_V} = V - {V_d}$. To ensure that the controller is not constrained by the initial velocity tracking error, an error transformation function ${\bar e_V} = \varphi {e_V}$ is introduced as defined by (5). Note that the initial value of $\varphi (t)$ is a small number $\beta$, which can converge from $\beta$ to $1$ within 
$T_P$. As long as the initial value $\beta$ is chosen sufficiently small, 
${\bar e_V}$ can definitely be kept within the prescribed performance function. From the subsequent proofs, it can be seen that $\beta$ affects the performance of fixed-time convergence; if $\beta$ is 0, the property of fixed-time convergence will disappear. To ensure the desired steady-state performance of the velocity tracking error, the prescribed performance transformation error is defined as:
\begin{equation}
	{\varepsilon _1} = \ln (\frac{{1 + {\xi _1}}}{{1 - {\xi _1}}}),
\end{equation}
where ${\xi _1} = \frac{{{{\bar e}_V}}}{{{\rho _1}}}$, $\rho _1$ is selected from reference \cite{label15}. To achieve prescribed performance control, consider the following Lyapunov candidate function:
\begin{equation}
	{V_1} = \frac{1}{2}\varepsilon _1^2.
\end{equation}

Take the derivative of $V_1$ and substitute (4) to obtain:
\begin{equation}
	\begin{aligned}
		\dot{V}_1 &= \varepsilon_1 \dot{\varepsilon}_1 \\
		&= \frac{2\varepsilon_1}{\rho_1(1 - \xi_1^2)}(\dot{\varphi} e_V + \varphi \dot{e}_V - \dot{\rho}_1 \frac{\bar{e}_V}{\rho_1}) \\
		&= \frac{2\varepsilon_1}{\rho_1(1 - \xi_1^2)} \left( \varphi (g_V \Phi_d + f_V + D_V - \dot{V}_d) - \dot{\rho}_1 \frac{\bar{e}_V}{\rho_1} \right) \\
		&+ \frac{2\varepsilon_1 \dot{\varphi} e_V}{\rho_1(1 - \xi_1^2)}.
	\end{aligned}
\end{equation}

The controller is designed in the following form:
\begin{equation}
\begin{aligned}
	{\Phi _d} &= ( - {f_V} - {\rho _1}(1 - \xi _1^2)({k_{v2}}{\varepsilon _1}^r + {k_{v3}}\tanh (\frac{{{\varepsilon _1}}}{{{l_{v2}}}}))\\
	 &\quad - {k_{v1}}{\varepsilon _1} - \frac{{{k_{v4}}{\varepsilon _1}\varphi _m^2e_v^2}}{{{\lambda _{v1}}{\rho _1}(1 - \xi _1^2)}} - \frac{{{\varepsilon _1}\varphi }}{{{\lambda _{v2}}{\rho _1}(1 - \xi _1^2)}}\\
	 &\quad - {{\hat D}_V} + {{\dot V}_d} + {{\dot \rho}_1}\frac{{{e_V}}}{{{\rho _1}}})/{g_v}.
\end{aligned}
\end{equation}

Let ${\dot \varphi _{\max }} = {\varphi _m},{\varphi _{\min }} = \beta $, substituting the controller (14) into (13) and using Young's inequality, we derive:
{\small
\begin{equation}
	\begin{aligned}
		\dot{V}_1 &\le -\frac{2\varepsilon_1^2\varphi_m^2e_v^2}{\lambda_{v1}\rho_1^2(1 - \xi_1^2)^2}(k_{v4}\beta - 1) + \frac{\lambda_{v1}}{2} + 2cl_{v1} + \frac{\lambda_{v2}E_V^2}{2} \\
		&\quad - \frac{2k_{v1}\varepsilon_1^2\varphi}{\rho_1(1 - \xi_1^2)} - 2\varphi(k_{v2}\varepsilon_1^{r + 1} + k_{v3}\varepsilon_1\tanh(\frac{\varepsilon_1}{l_{v2}})) \\
		&\le \!- 2\beta(k_{v2}\varepsilon_1^{r + 1} \!+k_{v3}\varepsilon_1\tanh(\frac{\varepsilon_1}{l_{v2}}))\!+ D_t\\
		&\le -k_{v2}\beta 2^{\frac{r + 3}{2}}V_1^{\frac{r + 1}{2}} - 2^{\frac{3}{2}}k_{v3}\beta V_1^{\frac{1}{2}} + D_1.
	\end{aligned}
\end{equation}
}
where ${D_1} = 2c{k_{v3}}{l_{v2}} +D_t,D_t= \frac{{{\lambda _{v1}}}}{2} + 2c{l_{v1}} + \frac{{{\lambda _{v2}}E_V^2}}{2}$.
\subsection{Altitude Controller Design:}
Adhering to the backstepping approach, the controller for the altitude subsystem (4) is structured into four distinct components. 

Define the error variable:
	${e_h} = h - {h_d}, 
	{e_\gamma} = \hat{\gamma}  - {x_{1d}}, 
	{e_\theta } = \theta  - {x_{2d}}, 
	{e_Q} = Q - {x_{3d}},$
where ${h_d}$ is the expected tracking signal, ${x_{1d}},{x_{2d}},{x_{3d}}$ are three new state variables to avoid differential explosion, which are defined as:
\begin{equation}
	{{\dot x}_{id}} = \frac{{ - {y_i}^r - \tanh (\frac{{{y_i}}}{{{l_{i}}}}) - {y_i}}}{{{\tau _i}}},
\end{equation}
where $i=1, 2, 3, l_i=l_{h2}, l_{r2}, l_{\theta 2}$. 
Define the filtering errors generated by the new state ${x_{1d}},{x_{2d}},{x_{3d}}: $
$
	{y_1} = {x_{1d}} - \bar \gamma, 
	{y_2} = {x_{2d}} - \bar \theta, 
	{y_3} = {x_{3d}} - \bar Q, 
$
where  $\bar \gamma, \overline \theta, \overline Q $ are the smooth virtual control signal to be designed later and $|\dot {\bar \gamma} | < {M_1}, |\dot {\bar \theta} | < {M_2}, |{\rm{\dot{ \bar Q}}}| < {M_3}$ hold where ${M_1}, {M_2}, {M_3}$ are the unknown positive constants. With these variables established, we can move forward with designing the controller:

\textbf{Step1.} Analogous to the velocity subsystem,  ${\bar e_h} = {\varphi _2}{e_h}$ is introduced. The prescribed performance transformation error is defined as follows:
\begin{equation}
	{\varepsilon _2} = \ln (\frac{{1 + {\xi _2}}}{{1 - {\xi _2}}}),
\end{equation}
where ${\xi _2} = \frac{{{{\bar e}_h}}}{{{\rho _2}}}$, $\rho _2$ is selected from reference \cite{label15}. If ${\varepsilon _2}$ is bounded, then ${\xi _2} \in ( - 1,1)$, and therefore ${\bar e_h}(t) < {\rho _2}(t)$ holds for all $t>0$. Consider the following Lyapunov candidate function:
\begin{equation}
	{V_{21}} = \frac{1}{2}\varepsilon _2^2.
\end{equation}

Differentiating $V_{21}$ and combining equations (4) with (17) yields:
\begin{equation}
	\begin{aligned}
		{{\dot V}_{21}} &= {\varepsilon _2}{{\dot \varepsilon }_2}\\
		 &= \frac{{2{\varepsilon _2}}}{{{\rho _2}(1 - \xi _2^2)}}({{\dot \varphi }_2}{e_h} + {\varphi _2}{{\dot e}_h} - {{\dot \rho }_2}\frac{{{{\bar e}_h}}}{{{\rho _2}}})\\
		& = \frac{{2{\varepsilon _2}{{\dot \varphi }_2}{e_h}}}{{{\rho _2}(1 - \xi _2^2)}} + \frac{{2{\varepsilon _2}}}{{{\rho _2}(1 - \xi _2^2)}}({\varphi _2}{g_h}(\hat \gamma  - {x_{1d}} + {x_{1d}} - \bar \gamma )\\
		 &\quad + {\varphi _2}{D_h} + {\varphi _2}{g_h}\bar \gamma  - {\varphi _2}{{\dot h}_d} - {{\dot \rho }_2}\frac{{{{\bar e}_h}}}{{{\rho _2}}}).
	\end{aligned}
\end{equation}

The virtual controller is designed in the following form:
\begin{equation}
  \begin{aligned}
	\bar \gamma  &= ({{\dot h}_d} + {{\dot \rho }_2}\frac{{{e_h}}}{{{\rho _2}}} - {k_{h1}}{\varepsilon _2}\\
	 &\quad - ({k_{h2}}\varepsilon _2^r + {k_{h3}}\tanh (\frac{{{\varepsilon _2}}}{{{l_{h1}}}})){\rho _2}(1 - \xi _2^2)\\
	 &\quad - \frac{{{\varepsilon _2}{\varphi _2}}}{{{\lambda _{h2}}{\rho _2}(1 - \xi _2^2)}} - \frac{{{\varepsilon _2}{\varphi _2}}}{{{\lambda _{h3}}{\rho _2}(1 - \xi _2^2)}}\\
	 &\quad - \frac{{{\varepsilon _2}{\varphi _2}}}{{{\lambda _{h4}}{\rho _2}(1 - \xi _2^2)}} - \frac{{{k_{h4}}{\varepsilon _2}\varphi _{2m}^2e_h^2}}{{{\lambda _{h1}}{\rho _2}(1 - \xi _2^2)}})/{g_h}.
  \end{aligned}
\end{equation}

Implementing the virtual control signal (20) in equation (19) while employing Young's inequality leads to:
{\small
\begin{equation}
	\begin{aligned}
		{\dot{V}_{21}} &\le  - \frac{{2\varepsilon _2^2\varphi _{2m}^2e_h^2}}{{{\lambda _{h1}}\rho _2^2{{(1 - \xi _2^2)}^2}}}({k_{h4}}{\beta _2} - 1) + \frac{{{\lambda _{h2}}g_h^2e_{\hat \gamma }^2}}{2}\\
		&\quad + \frac{{{\lambda _{h3}}g_h^2y_1^2}}{2} - \frac{{2{\varphi _2}{k_{h1}}\varepsilon _2^2}}{{{\rho _2}(1 - \xi _2^2)}} - 2{\beta _2}{k_{h2}}\varepsilon _2^{r + 1}\\
		&\quad - 2{\beta _2}{k_{h3}}{\varepsilon _2}\tanh (\frac{{{\varepsilon _2}}}{{{l_{h1}}}}) + \frac{{{\lambda _{h4}}E_h^2}}{2} + \frac{{{\lambda _{h1}}}}{2}\\
		& \le  - {2^{\frac{{r + 3}}{2}}}{\beta _2}{k_{h2}}V_{21}^{\frac{{r + 1}}{2}} - {2^{\frac{3}{2}}}{\beta _2}{k_{h3}}V_{21}^{\frac{1}{2}}\\
		&\quad + \frac{{{\lambda _{h2}}g_h^2e_{\hat \gamma }^2}}{2} + \frac{{{\lambda _{h3}}g_h^2y_1^2}}{2} + {D_{21}},
	\end{aligned}
\end{equation}
}
where ${D_{21}} = \frac{{{\lambda _{h1}}}}{2} + 2{\beta _2}{k_{h3}}c{l_{h1}} + \frac{{{\lambda _{h4}}E_h^2}}{2}$. Consider another Lyapunov candidate function:
\begin{equation}
	{V_{22}} = \frac{{y_1^2}}{2}.
\end{equation}
Through differentiation of $V_{22}$ in conjunction with equations (16), the following expression emerges:
{\small
\begin{equation}
	\begin{aligned}
		{{\dot V}_{22}} &\!\le \frac{{ \!- {y_1}^{r \!+ 1} \!- {y_1}\tanh (\frac{{{y_1}}}{{{l_{h2}}}}) \!- y_1^2}}{{{\tau _1}}} \!+ {M_1}|{y_1}|\\
		& \!\le \frac{{ \!- {2^{\frac{{r \!+ 1}}{2}}}V_{22}^{\frac{{r \!+ 1}}{2}}}}{{{\tau _1}}} - \frac{{{2^{\frac{1}{2}}}V_{22}^{\frac{1}{2}}}}{{{\tau _1}}} \!+ \frac{{c{l_{h2}}}}{{{\tau _1}}} \!+ \frac{{{\lambda _{h4}}}}{2} \!- (\frac{1}{{{\tau _1}}} \!- \frac{{M_1^2}}{{2{\lambda _{h4}}}})y_1^2\\
		& \!\le \frac{{ \!- {2^{\frac{{r \!+ 1}}{2}}}V_{22}^{\frac{{r \!+ 1}}{2}}}}{{{\tau _1}}} \!- \frac{{{2^{\frac{1}{2}}}V_{22}^{\frac{1}{2}}}}{{{\tau _1}}} \!- (\frac{1}{{{\tau _1}}} \!- \frac{{M_1^2}}{{2{\lambda _{h4}}}})y_1^2 \!+ {D_{22}},
	\end{aligned}
\end{equation}
}
where ${D_{22}} = \frac{{c{l_{h2}}}}{{{\tau _1}}} + \frac{{{\lambda _{h4}}}}{2}$.

\textbf{Step2.}  Differentiation of the flight path angle channel error produces:
\begin{equation}
	\begin{aligned}
		{{\dot e}_{\hat \gamma }} &= {g_\gamma }\theta  + {f_\gamma } + {D_{\hat \gamma }} - {{\dot x}_{1d}}\\
		& = {g_\gamma }{e_\theta } + {g_\gamma }{y_2} + {g_\gamma }\bar \theta  + {f_\gamma } + {D_{\hat \gamma }} - {{\dot x}_{1d}}.
	\end{aligned}
\end{equation}

The virtual control law is designed as follows:
\begin{equation}
	\bar \theta  \!= ( \!- {f_\gamma } \!- {\hat D_{\hat \gamma }} \!+ {\dot x_{1d}} \!- {k_{\hat \gamma 1}}{e_{\hat \gamma }} \!- {k_{\hat \gamma 2}}e_{\hat \gamma }^r \!- {k_{\hat \gamma 3}}\tanh (\frac{{{e_{\hat \gamma }}}}{{{l_{\hat \gamma 1}}}}))/{g_\gamma }.
\end{equation}

Consider the following candidate Lyapunov function:
\begin{equation}
	{V_3} = {V_{31}} + {V_{32}} = \frac{1}{2}e_{\hat \gamma }^2 + \frac{1}{2}y_2^2.
\end{equation}

The time derivative of Lyapunov function $V_3$ gives:
{\small
\begin{equation}
	\begin{aligned}
		{{\dot V}_3} &= {e_{\hat \gamma }}({g_\gamma }{e_\theta } + {g_\gamma }{y_2} + {E_{\hat \gamma }} - {k_{\hat \gamma 1}}{e_{\hat \gamma }} - {k_{\hat \gamma 2}}e_{\hat \gamma }^r - {k_{\hat \gamma 3}}\tanh (\frac{{{e_{\hat \gamma }}}}{{{l_{\hat \gamma 1}}}}))\\
		&\quad + {y_2}(\frac{{ - y_2^r - \tanh (\frac{{{y_2}}}{{{l_{\hat \gamma 2}}}}) - {y_2}}}{{{\tau _2}}}) + {M_2}|{y_2}|\\
		& \le  - {2^{\frac{{r + 1}}{2}}}{k_{\hat \gamma 2}}V_{31}^{\frac{{r + 1}}{2}} - {k_{\hat \gamma 3}}{2^{\frac{1}{2}}}V_{31}^{\frac{1}{2}} - \frac{1}{{{\tau _2}}}{2^{\frac{{r + 1}}{2}}}V_{32}^{\frac{{r + 1}}{2}}\\
		&\quad - \frac{1}{{{\tau _2}}}{2^{\frac{1}{2}}}V_{32}^{\frac{1}{2}} - ({k_{\hat \gamma 1}} - \frac{1}{{2{\lambda _{\hat \gamma 1}}}} - \frac{1}{{2{\lambda _{\hat \gamma 2}}}} - \frac{1}{{2{\lambda _{\hat \gamma 3}}}}){e_{\hat \gamma }}^2\\
		&\quad - (\frac{1}{{{\tau _2}}} - \frac{{{\lambda _{\hat \gamma 2}}{g_\gamma }^2}}{2} - \frac{{{\lambda _{\hat \gamma 4}}}}{2}){y_2}^2 + \frac{{{\lambda _{\hat \gamma 1}}{g_\gamma }^2{e_\theta }^2}}{2} + {D_3},
	\end{aligned}
\end{equation}
}
where ${D_3} = \frac{{{\lambda _{\hat \gamma 3}}{E_{\hat \gamma }}^2}}{2} + \frac{{{M_2}^2}}{{2{\lambda _{\hat \gamma 4}}}} + c{k_{\hat \gamma 3}}{l_{\hat \gamma 1}} + \frac{{c{l_{\hat \gamma 2}}}}{{{\tau _2}}}$.

\textbf{Step3.} Taking the derivative of the pitch angle channel error results in:
\begin{equation}
	\begin{aligned}
		{{\dot e}_\theta } &= Q + {D_\theta } - {{\dot x}_{2d}}\\
		& = {e_Q} + {y_3} + \bar Q + {D_\theta } - {{\dot x}_{2d}}.
	\end{aligned}
\end{equation}

The candidate Lyapunov function is designed as follows:
\begin{equation}
	{V_4} = {V_{41}} + {V_{42}} = \frac{1}{2}e_\theta ^2 + \frac{1}{2}y_3^2.
\end{equation}

By differentiating $V_4$, we obtain:
\begin{equation}
	\begin{aligned}
		{{\dot V}_4} &= {e_\theta }({e_Q} + {y_3} + \bar Q + {D_\theta } - {{\dot x}_{2d}})\\
		&\quad + {y_3}(\frac{{ - y_3^r - \tanh (\frac{{{y_3}}}{{{l_{\theta 2}}}}) - {y_3}}}{{{\tau _3}}}) + {M_3}|{y_3}|.
	\end{aligned}
\end{equation}

The virtual control law is designed as follows:
\begin{equation}
	\bar Q =  - {\hat D_\theta } + {\dot x_{2d}} - {k_{\theta 1}}{e_\theta } - {k_{\theta 2}}e_\theta ^r - {k_{\theta 3}}\tanh (\frac{{{e_\theta }}}{{{l_{\theta 1}}}}).
\end{equation}

Incorporating the virtual control law (31) into equation (30) establishes:
{\small
\begin{equation}
	\begin{aligned}
		{{\dot V}_4} &\le  - {2^{\frac{{r + 1}}{2}}}{k_{\theta 2}}V_{41}^{\frac{{r + 1}}{2}} - {2^{\frac{1}{2}}}{k_{\theta 3}}V_{41}^{\frac{1}{2}} - \frac{{{2^{\frac{{r + 1}}{2}}}}}{{{\tau _3}}}V_{42}^{\frac{{r + 1}}{2}} - \frac{{{2^{\frac{1}{2}}}}}{{{\tau _3}}}V_{42}^{\frac{1}{2}}\\
		&\quad - ({k_{\theta 1}} - \frac{1}{{2{\lambda _{\theta 1}}}} - \frac{1}{{2{\lambda _{\theta 2}}}} - \frac{1}{{2{\lambda _{\theta 3}}}} - \frac{1}{{2{\lambda _{\theta 4}}}}){e_\theta }^2\\
		&\quad - (\frac{1}{{{\tau _3}}} - \frac{{{\lambda _{\theta 5}}}}{2} - \frac{{{\lambda _{\theta 2}}}}{2}){y_3}^2 + \frac{{{\lambda _{\theta 1}}e_Q^2}}{2} + {D_4},
	\end{aligned}
\end{equation}
}
where ${D_4} = \frac{{{\lambda _{\theta 4}}\chi _2^2}}{2} + \frac{{{\lambda _{\theta 3}}E_\theta ^2}}{2} + \frac{{{M_3}^2}}{{2{\lambda _{\theta 5}}}} + c{k_{\theta 3}}{l_{\theta 1}} + \frac{{c{l_{\theta 2}}}}{{{\tau _3}}}$.

\textbf{Step4.} Differentiation of the pitch angle rate error generates:
\begin{equation}
	{\dot e_Q} = {g_Q}{\delta _{ed}} + {f_Q} + {D_Q} - {\dot x_{3d}}.
\end{equation}

The altitude subsystem controller is designed in the following form:
\begin{equation}
	{g_Q}{\delta _{ed}} \!=  \!- {k_{Q1}}{e_Q} \!- {f_Q} \!- {\hat D_Q} \!+ {\dot x_{3d}} \!- {k_{Q2}}e_Q^r \!- {k_{Q3}}\tanh (\frac{{{e_Q}}}{{{l_{Q3}}}}).
\end{equation}

The candidate Lyapunov function is designed as follows:
\begin{equation}
	{V_5} = \frac{1}{2}{e_Q}^2.
\end{equation}

Differentiating $V_5$ yields:
{\small
\begin{equation}
	\begin{aligned}
		{{\dot V}_5} &= {e_Q}{{\dot e}_Q}\\
		& = {e_Q}({E_Q} - {k_{Q1}}{e_Q} - {k_{Q2}}e_Q^r - {k_{Q3}}\tanh (\frac{{{e_Q}}}{{{l_{Q3}}}}))\\
		& \le c{l_{Q2}} + \frac{{{e_Q}^2}}{{2{\lambda _{Q1}}}} + \frac{{{\lambda _{Q1}}{E_Q}^2}}{2} - {k_{Q1}}{e_Q}^2 - {k_{Q2}}e_Q^{r + 1}\\
		&\quad - {k_{Q3}}{e_Q}\tanh (\frac{{{e_Q}}}{{{l_{Q3}}}})\\
		& \!\le  \!- ({k_{Q1}} \!- \frac{1}{{2{\lambda _{Q1}}}}){e_Q}^2 \!- {k_{Q2}}{2^{\frac{{r \!+ 1}}{2}}}V_5^{\frac{{r \!+ 1}}{2}} \!- {k_{Q3}}{2^{\frac{1}{2}}}V_5^{\frac{1}{2}} \!+ {D_5},
	\end{aligned}
\end{equation}
}
where ${D_5} = c{l_{Q2}} + \frac{{{\lambda _{Q1}}{E_Q}^2}}{2} + c{k_{Q3}}{l_{Q3}}$.
\subsection{Stability Analysis}
\noindent\textbf{Theorem 2.} Considering the closed-loop system with controllers (14) and (34), virtual control laws (20), (25) and (31), the observers (2), (3) and (6), and filtered states (16), the following results will hold:

1. The initial value of the prescribed function is independent of the initial errors of the velocity and altitude subsystems and the tracking errors will converge within the prescribed performance function within $T_p$.

2. The velocity and altitude tracking errors converge to a given value within an appointed time $T_s$.

3. All signals of the closed-loop system converge to a small neighborhood of the origin within a fixed time $T$.

\noindent\textbf{Proof:} Firstly, to prove the boundedness and practical fixed-time convergence properties of all signals in the closed-loop system, consider a Lyapunov candidate function:
\begin{equation}
	V = {V_1} + {V_2} + {V_3} + {V_4} + {V_5}.
\end{equation}

Through differentiation and substitution in equations (15), (21), (23), (27), (32), and (36), we arrive at:
{\small
\begin{equation}
	\begin{aligned}
		\dot V &= {{\dot V}_1} + {{\dot V}_2} + {{\dot V}_3} + {{\dot V}_4} + {{\dot V}_5}\\
		& \le  - {k_{v2}}\beta {2^{\frac{{r + 3}}{2}}}V_1^{\frac{{r + 1}}{2}} - {2^{\frac{{r + 3}}{2}}}{\beta _2}{k_{h2}}V_{21}^{\frac{{r + 1}}{2}} - \frac{{{2^{\frac{{r + 1}}{2}}}V_{22}^{\frac{{r + 1}}{2}}}}{{{\tau _1}}}\\
		&\quad - {2^{\frac{{r + 1}}{2}}}{k_{\hat \gamma 2}}V_{31}^{\frac{{r + 1}}{2}} - \frac{1}{{{\tau _2}}}{2^{\frac{{r + 1}}{2}}}V_{32}^{\frac{{r + 1}}{2}} - {2^{\frac{{r + 1}}{2}}}{k_{\theta 2}}V_{41}^{\frac{{r + 1}}{2}}\\
		&\quad \!-\! \frac{{{2^{\frac{{r + 1}}{2}}}}}{{{\tau _3}}}V_{42}^{\frac{{r + 1}}{2}} \!-\! {k_{Q2}}{2^{\frac{{r + 1}}{2}}}V_5^{\frac{{r + 1}}{2}} \!-\! {2^{\frac{3}{2}}}{k_{v3}}\beta V_1^{\frac{1}{2}} \!-\! {2^{\frac{3}{2}}}{\beta _2}{k_{h3}}V_{21}^{\frac{1}{2}}\\
		&\quad - \frac{{{2^{\frac{1}{2}}}V_{22}^{\frac{1}{2}}}}{{{\tau _1}}} - {k_{\hat \gamma 3}}{2^{\frac{1}{2}}}V_{31}^{\frac{1}{2}} - \frac{1}{{{\tau _2}}}{2^{\frac{1}{2}}}V_{32}^{\frac{1}{2}} - {2^{\frac{1}{2}}}{k_{\theta 3}}V_{41}^{\frac{1}{2}}\\
		&\quad - \frac{{{2^{\frac{1}{2}}}}}{{{\tau _3}}}V_{42}^{\frac{1}{2}} - {k_{Q3}}{2^{\frac{1}{2}}}V_5^{\frac{1}{2}} - (\frac{1}{{{\tau _1}}} - \frac{{M_1^2}}{{2{\lambda _{h4}}}} - \frac{{{\lambda _{h3}}g_h^2y_1^2}}{2})y_1^2\\
		&\quad - (\frac{1}{{{\tau _2}}} - \frac{{{\lambda _{\hat \gamma 2}}{g_\gamma }^2}}{2} - \frac{{{\lambda _{\hat \gamma 4}}}}{2}){y_2}^2 - (\frac{1}{{{\tau _3}}} - \frac{{{\lambda _{\theta 5}}}}{2} - \frac{{{\lambda _{\theta 2}}}}{2}){y_3}^2\\
		&\quad - ({k_{\hat \gamma 1}} - \frac{1}{{2{\lambda _{\hat \gamma 1}}}} - \frac{1}{{2{\lambda _{\hat \gamma 2}}}} - \frac{1}{{2{\lambda _{\hat \gamma 3}}}} - \frac{{{\lambda _{h2}}g_h^2}}{2}){e_{\hat \gamma }}^2\\
		&\quad - ({k_{\theta 1}} - \frac{1}{{2{\lambda _{\theta 1}}}} - \frac{1}{{2{\lambda _{\theta 2}}}} - \frac{1}{{2{\lambda _{\theta 3}}}} - \frac{1}{{2{\lambda _{\theta 4}}}} - \frac{{{\lambda _{\hat \gamma 1}}{g_{\hat \gamma }}^2}}{2}){e_\theta }^2\\
		&\quad - ({k_{Q1}} - \frac{1}{{2{\lambda _{Q1}}}} - \frac{{{\lambda _{\theta 1}}e_Q^2}}{2}){e_Q}^2 + D,
	\end{aligned}
\end{equation}
}
where $D = {D_1} + {D_{22}} + {D_{21}} + {D_3} + {D_4} + {D_5}$.With proper parameter selection and Lemma 1, we derive:
\begin{equation}
	\dot V \le  - {p_1}{V^{{a_1}}} - {p_2}{V^{{a_2}}} + D,
\end{equation}
where:${p_1} \!=\! {2^{2 - r}}\min (2{k_{v2}}\beta ,2{\beta _2}{k_{h2}},\frac{1}{{{\tau _1}}},{k_{\hat \gamma 2}},\frac{1}{{{\tau _2}}},{k_{\theta 2}},\frac{1}{{{\tau _3}}},{k_{Q2}})\\
	,{p_2} = {2^{\frac{1}{2}}}\min (2{k_{v3}}\beta ,2{\beta _2}{k_{h3}},\frac{1}{{{\tau _1}}},{k_{\hat \gamma 3}},\frac{1}{{{\tau _2}}},{k_{\theta 3}},\frac{1}{{{\tau _3}}},{k_{Q3}}),\\
	{a_1} = \frac{{r + 1}}{2},{a_2} = \frac{1}{2},r \in (1, + \infty ),
r$ is an odd integer.

Define the error vector $X = [{\varepsilon _1},{\varepsilon _2},{e_r},{e_\theta },{e_Q},$
${y_1},{y_2},{y_3}]$. The error vector will converge to a small neighborhood around the origin within a fixed time:
{\small
\begin{equation}
	{T_1} \le \frac{1}{{{p_2}w(1 - {a_2})}} + \frac{1}{{{p_1}w({a_1} - 1)}},
\end{equation}
}
{\small
\begin{equation}
	\{ \Omega :V(X) \le \min \{ {(\frac{D}{{(1 - w){p_1}}})^{\frac{1}{{{a_1}}}}},{(\frac{D}{{(1 - w){p_2}}})^{\frac{1}{{{a_2}}}}}\} \} .
\end{equation}
}

Therefore, $|{\bar e_i}| < {\rho _i},i = V,h$ holds for all $t>0$. According to the definition of the transformed error function, after time $T_p$, we have ${e_i} = {\bar e_i}$. Since ${T_s} > {T_p}$ in the design, it follows that after $T_p$, $|{e_i}| < {\rho _i},i = V,h$. Subsequently, the tracking errors of the closed-loop system will converge to the appointed values within the time $T_s$. Thus, the initial value of $e_i$ can be greater than the initial value of the prescribed performance function. Within time $T_p$ will converge inside the bounds of the prescribed performance function and remain within these bounds, converging to the appointed values within $T_s$. This completes the proof.
\section{Simulation Results}
Simulate the high-altitude acceleration cruise of a hypersonic vehicle, where the velocity command increases from 7846.4 ft/s to 10032 ft/s, and the altitude command climbs from 85,000 ft to 105,583 ft. The aerodynamic parameter uncertainties are set to 30\%, and system disturbances are selected according to \cite{label8}. The initial altitude error is 40 ft, and the initial velocity error is 8 ft/s. At $t=50$, an actuator fault occurs in the velocity subsystem with ${\lambda _\Phi } = 0.8,{f_\Phi } = 0.03$, and simultaneously, an actuator fault occurs in the altitude subsystem with ${\lambda _\delta } = 0.8,{f_\delta } = 0.05$. The relevant controller parameters are chosen as follows:
$
	{k_{v1}} = 3,{k_{v2}} = 2,{k_{v3}} = 1,{k_{v4}} = 10,{\lambda _{V1}} = {\lambda _{V2}} = 1,{T_{pV}} = 2.5,{\xi _{av}} = 6,{\xi _{bv}} = 0.2,{T_{sV}} = 10,\
	{k_{h1}} = 1.7,{k_{h2}} = {k_{h3}} = 1,{k_{h4}} = 8,{\lambda _{h1}} = {\lambda _{h2}} = {\lambda _{h3}} = {\lambda _{h4}} = 1,{T_{ph}} = 5,{\xi _{ah}} = 40.6,
	{\xi _{bh}} = 0.6,{T_{sh}} = 30,{k_{\hat \gamma 1}} = 0.5,{k_{\hat \gamma 2}} = {k_{\hat \gamma 3}} = 0.1,{k_{\theta 1}} = 1,{k_{\theta 2}} = {k_{\theta 3}} = 0.1,{k_{Q1}} = 2,
	{k_{Q2}} = {k_{Q3}} = 0.5,r = 3,{\tau _i} = 0.2,{l_i} = 0.1.$
The parameters selection for the state reconstruction observer are: ${d_h} = 20,d = 15,{\eta _0} = {\eta _1} = {\eta _2} = {\eta _3} = 1.5$. The parameters selection for the lumped disturbance observer are:
$
	{l_{V1}} = {l_{V2}} = 5,{l_{V3}} = 1,{\Gamma _{wV}} = 1.2,
	{l_{h1}} = {l_{h2}} = 5,{l_{h3}} = 1,{\Gamma _{wh}} = 1.5,
	{l_{\hat \gamma 1}} = {l_{\hat \gamma 2}} = 10,{l_{\hat \gamma 3}} = 1,{\Gamma _{w\hat \gamma }} = 2,
	{l_{\theta 1}} = {l_{\theta 2}} = 10,{l_{\theta 3}} = 1,{\Gamma _{w\theta }} = 2,
	{l_{Q1}} = {l_{Q2}} = 20,{l_{Q3}} = 1,{\Gamma _{wQ}} = 10,
	{x_V} = V,{x_h} = {[V,\hat \gamma ]^T},{x_{\hat \gamma }} = {[V,\hat \gamma ,\theta ]^T},{x_\theta } = {[V,\hat \gamma ,\theta ]^T},{x_Q} = {[V,\hat \gamma ,\theta ,Q]^T},
	{\alpha _i} = 0.5,{\beta _i} = 2$,$V \in [7500{\rm{ft/s,}}11000{\rm{ft/s}}],\hat \gamma  \in [ - 2\deg ,2\deg ],\theta  \in [ - 5\deg ,5\deg ],Q \in [ - 10\deg {\rm{/s}},10\deg {\rm{/s}}],$ the center of the neural networks are uniformly selected within the range of each state.
	
The simulation is conducted in two parts. The first part uses the algorithm designed in this paper to achieve tracking of the command signal. To verify the superiority of the designed algorithm, the second part selects the algorithm from reference \cite{label8} for comparison. The results of the first part of the simulation are shown in the figures below. Figure 1 demonstrates the reconstruction effect for $\alpha ,\gamma $:

\begin{figure}[!htb]
	\centering
	\includegraphics[width=\hsize]{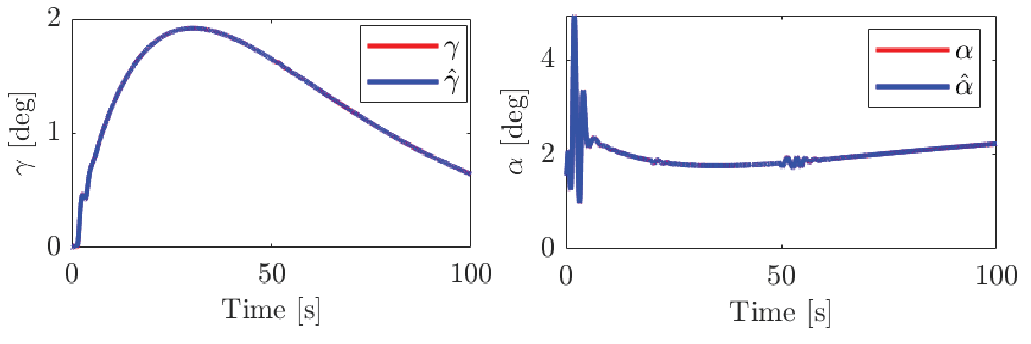}
	\caption{The reconstruction of unmeasurable state}
	\label{fig1}
\end{figure}
It can be observed that the designed observer successfully reconstructs the state $\alpha, \gamma $ effectively. Furthermore, Figure 2 shows that the designed controller can achieve good command tracking for a hypersonic aircraft with disturbances and actuator faults. Figures 3 to 4 show that even if the initial value of the tracking error is greater than the initial value of the prescribed performance function, the tracking error can quickly converge within the prescribed performance function and remain within it. Even in the event of actuator failures, effective control can still be achieved.

\begin{figure}[!htb]
	\centering
	\includegraphics[width=\hsize]{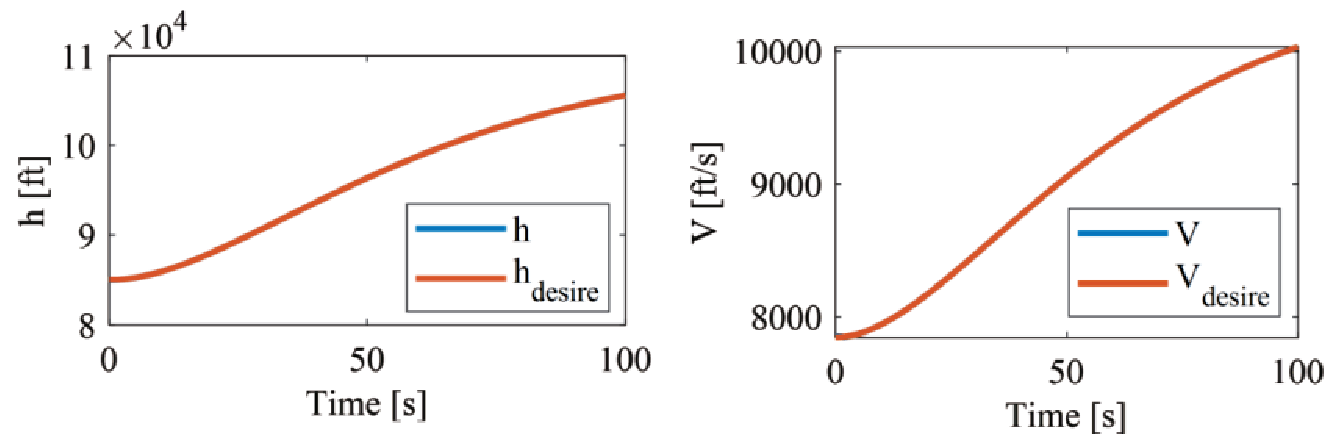}
	\caption{The tracking of command signals}
	\label{fig1}
\end{figure}
\begin{figure}[!htb]
	\centering
	\includegraphics[width=\hsize]{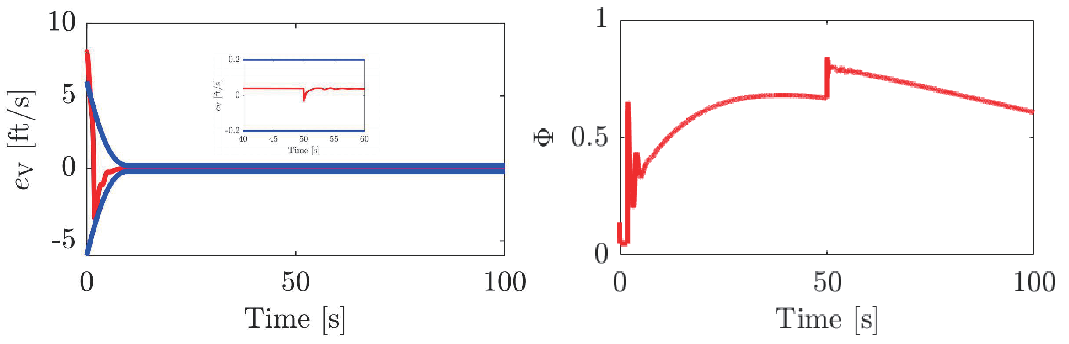}
	\caption{Velocity subsystem tracking error and control input}
	\label{fig1}
\end{figure}
\begin{figure}[!htb]
	\centering
	\includegraphics[width=\hsize]{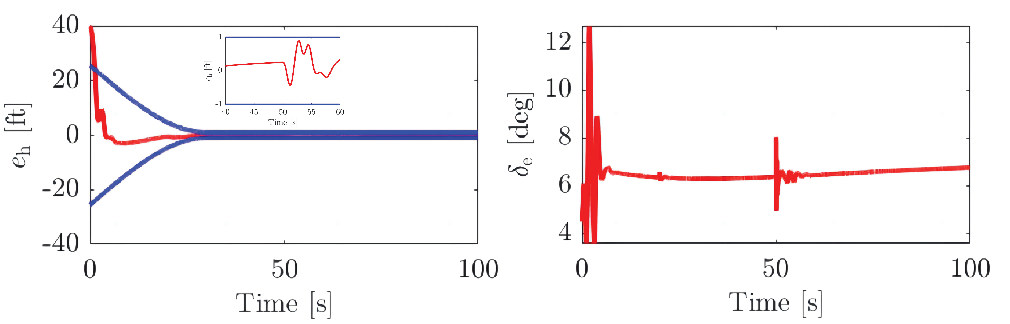}
	\caption{Altitude subsystem tracking error and control input}
	\label{fig1}
\end{figure}
The simulation results for the second part are shown in Figures 5 to 6. Since the initial error in reference [5] is not allowed to exceed the initial value of the prescribed performance function, the initial error was chosen to be close to the initial error of the prescribed performance function. To ensure a fair comparison, the parameters of both controllers were tuned. It can be observed that the newly designed controller represented by the brown line is less sensitive to initial errors compared to the controller in [5] represented by the blue line. And the proposed controller has a faster convergence speed, smaller overshoot, and is smoother in operation.
\begin{figure}[!htb]
	\centering
	\includegraphics[width=\hsize]{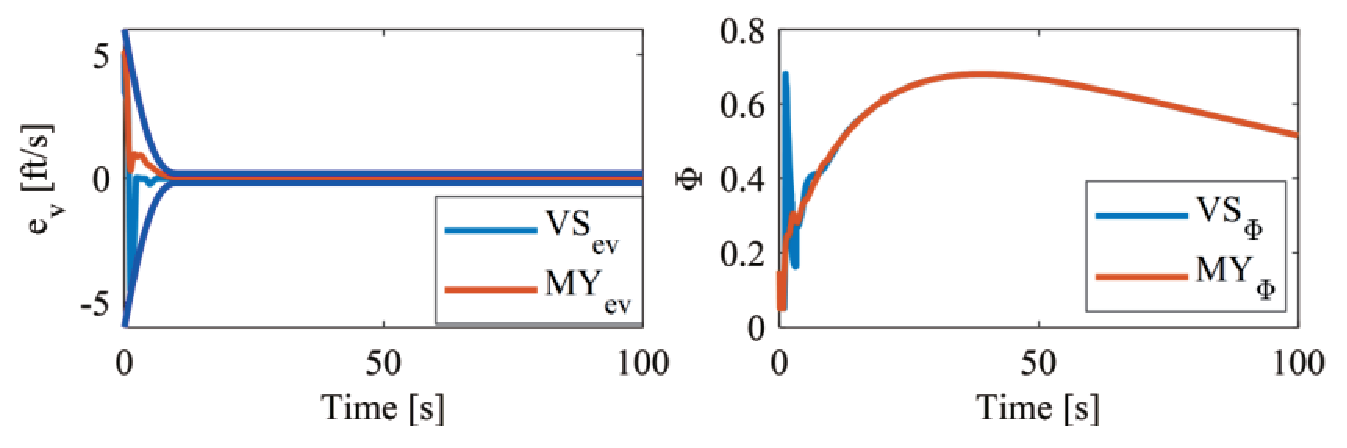}
	\caption{Comparison of velocity subsystem tracking performances}
	\label{fig1}
\end{figure}
\begin{figure}[!htb]
	\centering
	\includegraphics[width=\hsize]{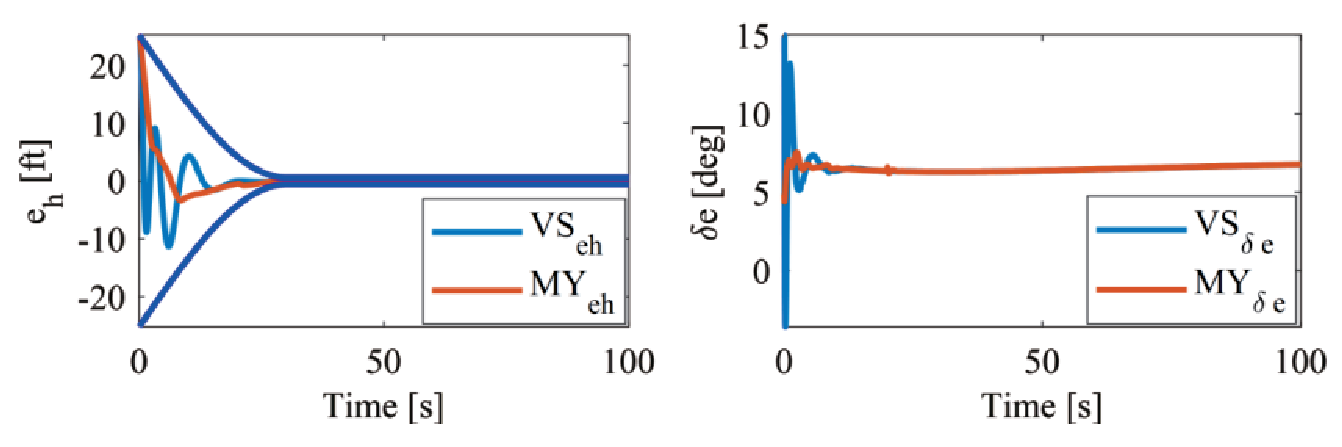}
	\caption{Comparison of altitude subsystem tracking performances}
	\label{fig1}
\end{figure}

\section{Conclusion}
\vspace{-0.2cm}
In this paper, for a hypersonic aircraft with unmeasurable states and actuator failures, a new type of state observers are constructed to reconstruct the unmeasurable states. Based on this, a novel fault-tolerant controller that does not depend on the initial state and has prescribed performance is designed to realize the tracking of command velocity and altitude trajectory signals. In addition, the proposed control strategy can ensure practical fixed-time convergence of the closed-loop system, thereby greatly enhancing the transient performance.

\end{document}